\documentclass[aps,10pt,prl,twocolumn,showpacs,superscriptaddress]{revtex4-1}
\usepackage{graphicx}

\begin{document}

\title{Differential Scattering Cross-Sections for the Different
  Product Vibrational States in the Ion-Molecule Reaction Ar$^+$ +
  N$_2$}

\author{S. Trippel}
\affiliation{Center for Free-Electron Laser Science, DESY, Notke-Stra{\ss}e 85, 22706 Hamburg, Germany}

\author{M. Stei}

\affiliation{Institut f{\"u}r Ionenphysik und Angewandte Physik,
  Universit{\"a}t Innsbruck, Technikerstra{\ss}e 25, 6020 Innsbruck,
  Austria}

\author{J. A. Cox}
\affiliation{Department of Plant Sciences, University of Cambridge, Downing Street, Cambridge, CB2 3EA, United Kingdom}

\author{R. Wester}
\email[]{E-mail: roland.wester@uibk.ac.at}
\affiliation{Institut f{\"u}r Ionenphysik und Angewandte Physik,
  Universit{\"a}t Innsbruck, Technikerstra{\ss}e 25, 6020 Innsbruck,
  Austria}

\date{\today}

%%%%%%%%%%%%%%%%%%%%%%%%%%%%%%%%%%%%%%%%%%%%%%%%%%%%%%%%%%%%%%%%%%%%%%%%%%%%%%

\begin{abstract}
  The charge transfer reaction Ar$^+$ + N$_2$ $\rightarrow$ Ar +
  N$_2^+$ has been investigated in a crossed beam experiment in
  combination with three-dimensional velocity map imaging. Angular
  differential state-to-state cross sections were determined as a
  function of the collision energy. We found that scattering into the
  first excited vibrational level dominates as expected, but only for
  scattering in the forward direction. Higher vibrational excitations
  up to $v'=6$ have been observed for larger scattering angles. For
  decreasing collision energy, scattering into higher scattering
  angles becomes increasingly important for all kinematically allowed
  quantum states. Our detailed measurements indicate that a
  quantitative agreement between experiment and theory for this basic
  ion-molecule reaction now comes within reach.
\end{abstract}

%%%%%%%%%%%%%%%%%%%%%%%%%%%%%%%%%%%%%%%%%%%%%%%%%%%%%%%%%%%%%%%%%%%%%%%%%%%%%%

\maketitle

%%%%%%%%%%%%%%%%%%%%%%%%%%%%%%%%%%%%%%%%%%%%%%%%%%%%%%%%%%%%%%%%%%%%%%%%%%%%%%

%electron transfer reactions
Gas phase studies of ion-molecule reactions have provided insight into
a multitude of chemical processes in environments where ions and
neutrals coexist. Ion-molecule reactions determine the abundance of
many of the complex species that can be detected in interstellar
molecular clouds and in planetary atmospheres
\cite{waite2007:sci,larsson2012:rpp}. The conceptually simple charge
transfer reactions are particularly interesting, as they were found to
explain the X-ray emission from comets
\cite{cravens2002:sci,lisse2001:sci} and may serve as a possible
acceleration mechanism of cosmic rays due to strong shock waves in
supernova remnants \cite{helder2009:sci}. Charge transfer is also of
cosmological importance in that it shapes the hydrogen chemistry in
the early universe \cite{savin2004:apj}. Laboratory measurements and
theoretical calculations are needed to provide the basis for modeling
these processes. In addition charge transfer reactions lead to
characteristic light emission from excited states that are useful to
determine the parameters of laboratory or technical plasmas, such as
temperature, velocity, electron density and charge states of ions
\cite{Perez2001,Rosmej2006}. To reach very low collision energies,
charge transfer has been studied near threshold in half collisions
\cite{wells2001:prl}. Recently, ultracold charge transfer reactions
have become of interest in studies of cold atom-ion collisions, which
are carried out to investigate quantum mechanical phenomena in
scattering processes at very low energies
\cite{zipkes2010:prl,schmid2010:prl,hall2012:prl}.

%how do the reactions evolve (hand-waving arguments)
Charge transfer reactions in gas phase evolve in many cases not on a
single potential energy surface and are therefore often accompanied by
a breakdown of the Born-Oppenheimer approximation. Since the shape of
the molecule changes upon charge transfer, state to state electron
transfer rates are often controlled by intra-molecular vibrational
motion \cite{Kleyn1980,tosi1991:prl,Huang2000:sci}. While in many cases the outcome
of a reaction at high collision energies is well described by the
properties of the isolated molecule the results at low energies are
not explained by a Frank-Condon treatment
\cite{candori2001:jcp,Huewel1984:jcp}. This behavior may be explained
by molecular bond distortions due to the electric field of the
incoming ion \cite{Lipeles1969:jcp} and to short range repulsive
interactions between the projectile and the target \cite{Kelley1977}.

%The Ar+ + N2 system and experiments
A model system of a gas phase non-Born-Oppenheimer reaction mechanism
is the charge transfer reaction
\begin{displaymath}
{\rm Ar}^+(^{2}{\rm P}_J) + {\rm N}_2(^{1}\Sigma_g^{+},v=0) 
\rightarrow
{\rm Ar}(^{1}{\rm S}_0) + {\rm N}_2^+ (^{2}\Sigma_g^{+},v').
\label{arn2:reaction}
\end{displaymath}
Despite its overall exoergicity the most abundant product channel at
low collision energy is N$_2^+$ in the first vibrationally excited
level, which is endoergic by 0.092\,eV. Experimental evidence was
obtained early on that this channel is formed despite having a much
lower Franck-Condon factor than the exoergic ground state
\cite{Lindinger1981:pra,Smith1981}. Later, angle differential cross
sections and vibrational and rotational state distribution have been
determined in the energy range of 0.3-3\,eV
\cite{Huewel1984:jcp,Rockwood1985:cpl,liao1986:jcp,Futrell1987,Birkinshaw1987:cp,Sonnenfroh1989}. Here,
crossed-beam experiments \cite{Birkinshaw1987:cp} found much higher
vibrational excitation in the collision energy range near 1\,eV
collision energy than radiofrequency ion guide experiments
\cite{liao1986:jcp}. Even after reconsidering the finite velocity and
angular resolution and the involved unfolding scheme of the
crossed-beam experiments this disagreement remained unresolved
\cite{howard1991:cpl}. Also state-of-the-art theoretical calculations
could not achieve an agreement with the measured product energy and
angular distributions
\cite{Spalburg1985,Parlant1987,nikitin1987,Clary1989:jcp}. More
recently, a semi-classical Landau-Zener model has been employed to
derive state-to-state cross sections
\cite{candori2001:jcp,candori2003:ijm}. In this model higher
vibrational excitations are predicted for larger scattering
angles. When we developed the crossed-beam spectrometer for
ion-molecule reactions based on the velocity map imaging technique
\cite{eppink1997:rsi}, we revisited the reaction of Ar$^+$ with N$_2$
and found again significant vibrational excitation, but still with
insufficient resolution to provide a clear picture of the reaction
\cite{mikosch2006:pccp}.

%why measured once again
In this letter we present detailed angular-differential scattering
cross sections together with a product vibrational state analysis for
the Ar$^+$ + N$_2$ charge transfer reaction. Using an optimized design
of the velocity map imaging potentials with respect to stray electric
fields and a full three-dimensional measurement of the product
velocity vectors, we are now able to assign the different vibrational
states of the N$_2⁺$ product ions in the measured kinetic energy
distributions. Our results clearly disagree with the previous
crossed-beam cross section measurements \cite{Birkinshaw1987:cp},
possibly caused by the difficulty to unambiguously unfold the previous
angle-resolved scattering data. Our measured angular distributions
change systematically with product vibrational state, qualitatively in
agreement with semi-classical calculations
\cite{candori2001:jcp,candori2003:ijm}.

%ion production
The experiment is based on our previous work on negative ion-molecule
reactions \cite{mikosch2006:pccp,mikosch2008:sci}. Crossed-beam
imaging studies of ion-molecule reactions have recently also been
reported for reactions of C$^+$ with NH$_3$ \cite{pei2012:jcp}. Here,
Ar$^+$ ions are produced by a combination of an electron gun and a
supersonic Ar gas pulse provided by a piezo-electric valve. In the
extraction volume of a Wiley McLaren mass spectrometer the ions are
extracted perpendicularly and accelerated towards the crossed-beam
spectrometer. Prior to collision the ions are decelerated inside a
cylindrical electrode just outside the spectrometer to the desired
kinetic energies in the range of 1.0-5.6\,eV, which are measured with
the velocity map imaging spectrometer (FWHM of about 200\,meV). Ar$^+$
may generally be produced in the P$_{1/2}$ and P$_{3/2}$ states by
electron impact ionization. As discussed below, we estimate a
contribution of less than 20\% of the P$_{1/2}$ state in this
experiment. The N$_2$ target beam is generated in a pulsed supersonic
expansion, provided by a piezo-electric valve with a stagnation
pressure of 2\,bar at a temperature of 70$^{\circ}$C. The central part
of the supersonic beam enters the scattering center after passing a
skimmer with an orifice diameter of 200\,$\mu$m, placed 35\,mm behind
the nozzle. For N$_2$ we expect translational and rotational
temperatures of about 5\,K \cite{trippel2009:jpb}. The laboratory
velocity of the molecules is measured after electron impact ionization
with the velocity map imaging setup to be around 830\,m/s, in good
agreement with expectation.

% VMI and ion detection
The ion and the molecular beam cross each other in the center of a
velocity map imaging stack at a scattering angle of 61$^\circ$. Once
the two reactant pulses have crossed, the velocity map imaging
electrodes are switched on to map the ion velocities onto the imaging
detector, which consists of a microchannel plate combined with a
phosphor screen and a CCD camera. The ion time-of-flight is obtained
by a photo-multiplier tube in combination with a time-to-digital
converter picking-up the rising edge of the light spot on the phosphor
screen.  Typical event rates are less than one ion per bunch crossing
with a background rate about two orders of magnitude lower. Images
with two or more detected ions were neglected to be able to correlate
transverse position and time-of-flight.  Ion-impact positions and
time-of-flight information are used to determine the three dimensional
velocity vector of the ions in the interaction region. Note that the
velocity vector is measured irrespective of the scattering angle,
yielding an effective $4\pi$ angular acceptance. In order to suppress
signals from the Ar$^+$ reactant beam, the detector is activated for a
period of 1\,$\mu s$ around the N$_2^+$ arrival time.

%%%%%%%%%%%%%%%%%%%%%%%%%%%%%%%%%%%%%%%%%%%%%%%%%%%%%%%%%%%%%%%%%%%%%%%%%%%%%%
% experimental results

\begin{figure}[htp]
  \begin{center}
    \includegraphics[width=1.0\columnwidth, trim=10cm 15cm 2cm 10cm,
      clip]{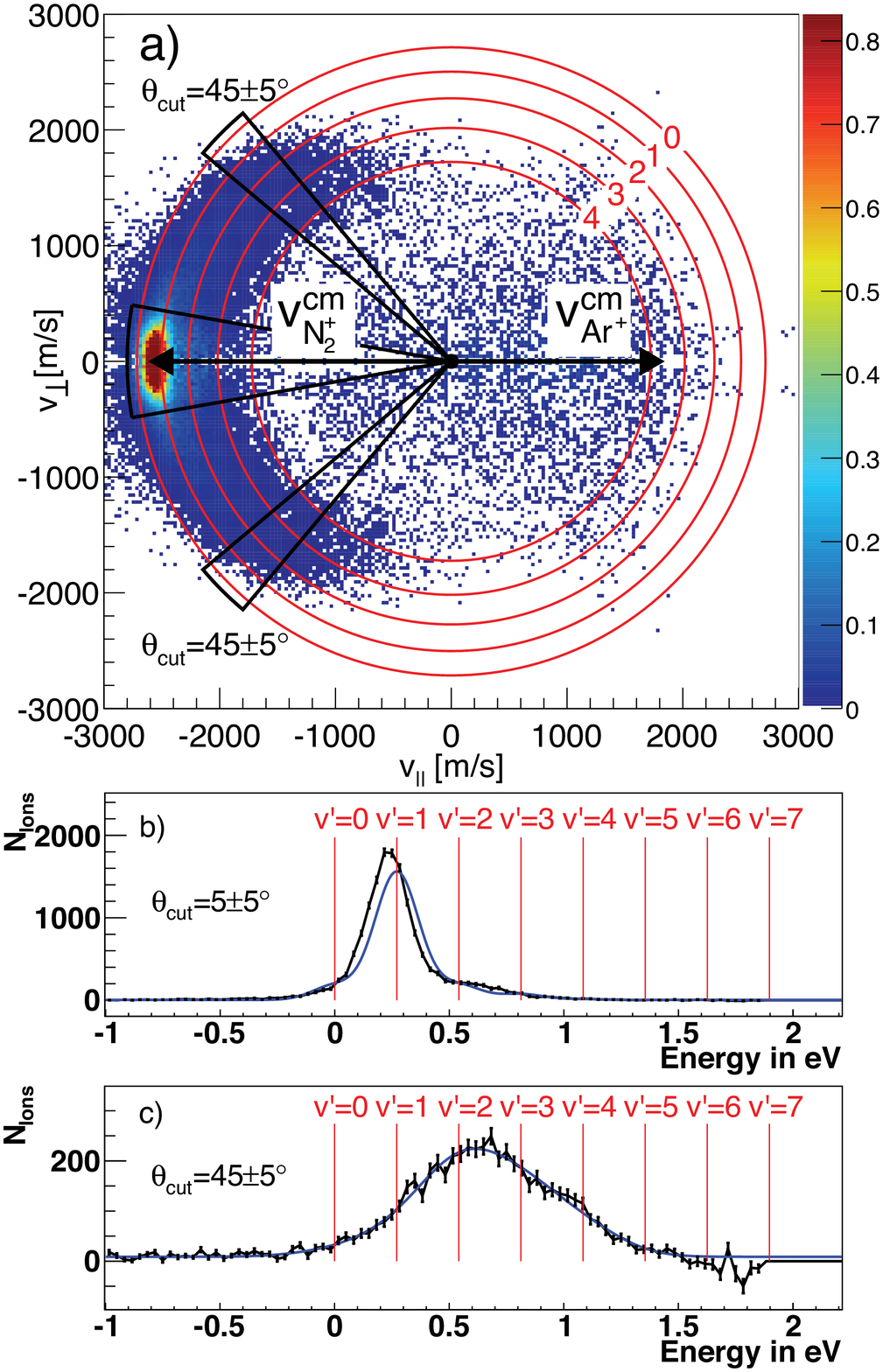}
    \caption{a) Differential scattering cross section for the reactive
      scattering of Ar$^+$+N$_2$ at a relative energy of 1.7\,eV. The
      image is obtained by rotating and weighting the measured three
      dimensional velocity distribution of N$_2^+$ ions. Reactant
      velocities are indicated by black arrows (N$_2$ along the
      negative and Ar$^+$ along the positive v$_{||}$ axis). Newton
      spheres are plotted which correspond to N$_2^+$ vibrational
      excitation with $v'=0-4$. Black lines indicate cuts on the
      scattering angle $\theta$. b) N$_2^+$ energy distribution with a
      cut on the scattering angle $\theta$=5$\pm5^\circ$. c) N$_2^+$
      energy distribution with a cut on the scattering angle
      $\theta$=45$\pm5^\circ$. Blue lines in panels b,c are fits to a
      sum of Gaussian, each centered at $v'=0-7$.}
    \label{2ddist.fig}
  \end{center}
\end{figure}

%the 2D histogram
We have measured the differential cross section of reaction
\ref{arn2:reaction} for relative collision energies of 0.3, 0.5, 0.8,
1.2, 1.7 and 2\,eV. A representative histogram of the cross section at
1.7\,eV is shown in Fig.\ \ref{2ddist.fig}a. It is obtained by
determining the three-dimensional velocity vector for each N$_2⁺$
product ion in the center-of-mass frame and computing its components
in the scattering plane. In order to compare our results with slice
and projected images used in previous measurements each velocity entry
is weighed by 1/$v_\perp$ where $v_\perp$ is the velocity component
perpendicular to the symmetry axis, i.\ e.\ the relative velocity axis
of the scattering process. For the same reason we also extend the
image by mirroring the data to negative values of $v_\perp$. The Ar$^+$
reactant velocity is indicated by an arrow pointing along the positive
v$_{||}$ axis and the N$_2$ educt velocity is indicated by an arrow
pointing along the negative v$_{||}$ axis. Newton rings mark the
product velocities corresponding to $v'=0-4$ vibrational quanta in the
N$_2⁺$ product ion. A sharp peak in the forward direction is observed
in the differential cross section. Additionally, a contribution of
larger scattering angles in the forward hemisphere is found. The
forward scattering peak shows product velocities mainly corresponding
to $v'=1$. The sideways scattered products clearly show vibrational
excitation in higher $v'$ states.

%scattering into selected scattering angle windows
For a quantitative analysis the angular distributions of each product
vibrational state are extracted from the measured cross section
images. Fig.\ \ref{2ddist.fig}b shows the internal energy distribution
of the product N$_2^+$ ions scattered into the narrow cone of forward
scattering angles $\theta=5\pm5^\circ$ (marked in black in
Fig.\ \ref{2ddist.fig}a). Red lines indicate the internal energy of
N$_2^+$ in the vibrational levels $v'=1-7$. As expected, most of the
product ions are excited with one vibrational quantum in agreement
with previous measurements and theoretical predictions. There is only
a small contribution of the levels with $v'=0$, 2 and 3. In addition
we estimate from the shown distribution a contribution from reactions
of the P$_{1/2}$ state of Ar$^+$ of less than 20\%. Otherwise a strong
peak at 0.36\,eV internal energy would have to be present in the
distribution. Such a small contribution is in accord with a
statistical mixture of P$_{3/2}$:P$_{1/2}$ of 2:1 and the lower
reactivity of Ar$^+$(P$_{1/2}$) \cite{kato1982:jcp}.  

The internal energy distribution of product ions scattered into
$\theta$=45$\pm5^\circ$ (also marked in black in
Fig.\ \ref{2ddist.fig}a) is shown in Fig.\,\ref{2ddist.fig}c. For
these scattering angles $v'=1$ is not the dominant product channel
anymore. Instead, the internal energy distribution exhibits higher
vibrational excitations with the majority of ions being scattered into
$v'=2$. It can be seen from Figs.\ \ref{2ddist.fig}b,c that the
individual product vibrational levels are not resolved. This can be
attributed to the experimental energy resolution dominated by the
finite energy width of the slow ion beam. The expected resolution
depends on the product scattering angle and is calculated to change
from 0.06\,eV for forward scattering to 0.2\,eV for backward
scattering. A second contribution to the measured broadening stems
from rotational excitation of N$_2^+$, which is on the order of
0.06\,eV \cite{Huewel1984:jcp,Sonnenfroh1989}.

\begin{figure}[tb]
  \begin{center}
    \includegraphics[width=1.0\columnwidth, trim=9cm 14cm 2.5cm 0cm, clip]{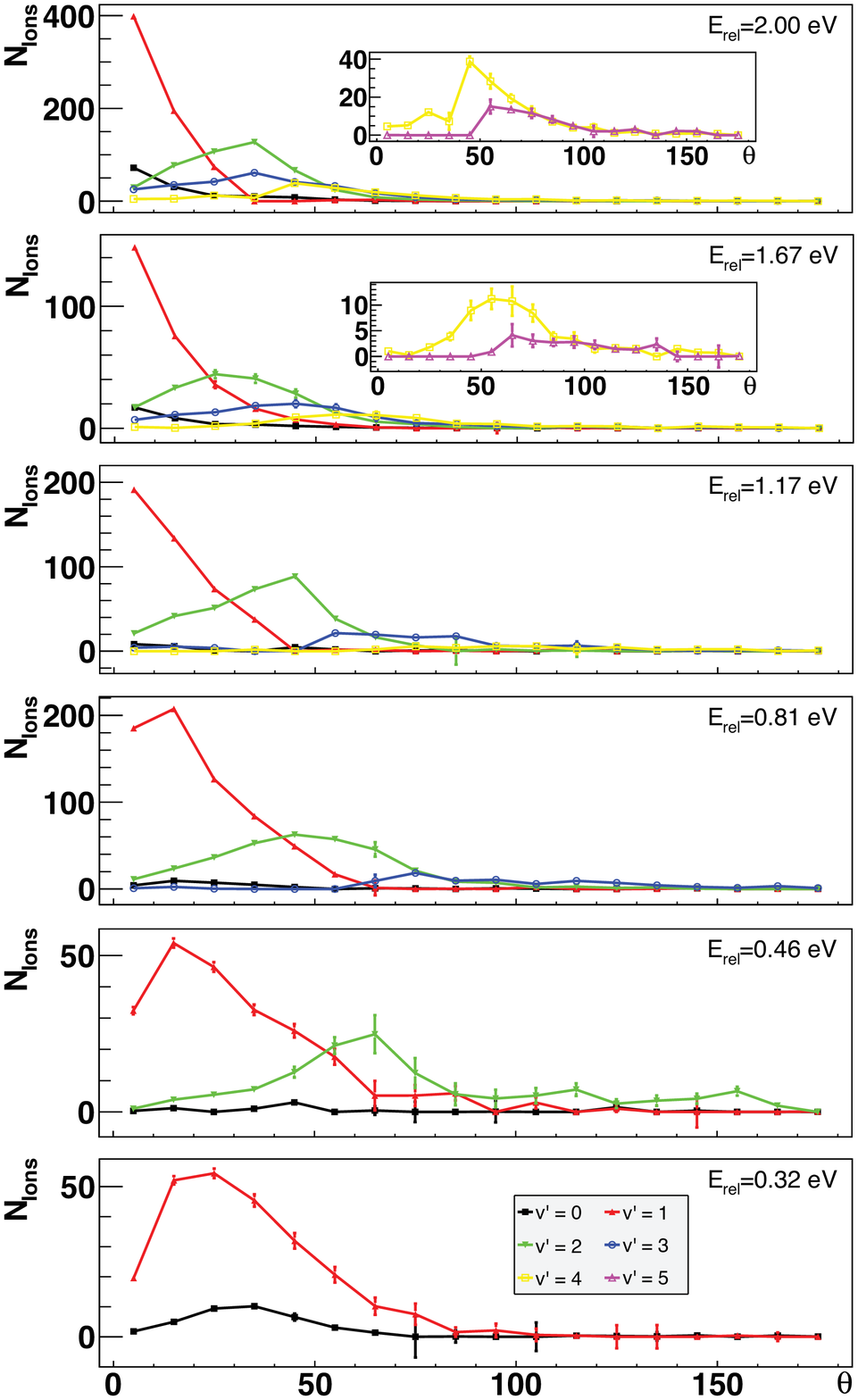}
    \caption{Number of ions scattered into vibrational state $v'=0-6$
      of the N$_2^+$ product ion as a function of scattering angle
      $\theta$ for all measured energies between 0.3\,eV and
      2\,eV. Scattering into $v'=1$ dominates for small scattering
      angles whereas higher vibrational excitation is more pronounced
      at larger angles. All kinematically allowed product levels are
      scattered into increasing larger angles with decreasing
      collisional energy.}
    \label{vib_dist.fig}
  \end{center}
\end{figure}

%how to get the vibrational distribution
To analyze the angular dependence of the scattering cross section for
each vibrational level separately, the relative vibrational
populations are extracted from the measured energy distributions for
10$^\circ$ scattering angle intervals between 0 and 180$^\circ$. For
this we fit a sum of Gaussian functions, each representing a different
vibrational level $v'$, to the energy distributions for each
scattering angle interval. Only the heights of the Gaussian functions
are free fit parameters, the mean of each Gaussian is pre-determined
by the vibrational excitation energy of each level $v'$. The Gaussian
widths represent the product energy resolution and are calculated from
the measured energy resolution of the reactant beams. The width is
assumed to be the same for all Gaussians in a single internal energy
distribution, but it changes as a function of the scattering angle, as
mentioned above. Finally, all Gaussians are multiplied with a step
function with its edge at the kinematical cutoff. For the intervals in
Figs.\ \ref{2ddist.fig}b,c, the fits of the sum of Gaussians are shown
as blue lines. The fitted heights and their standard deviation
accuracies are used to determine the number of N$_2^+$ product ions
scattered into a given vibrational level.

%relative vibrational excitation
Fig. \ref{vib_dist.fig} shows the number of ions in the vibrational
states $v'=1-6$ as a function of the scattering angle $\theta$ for all
measured collision energies between 0.3\,eV and 2\,eV. For all
energies vibrational excitation in the $v'=1$ channel dominates in
forward direction whereas higher vibrational excitation is more
prominent at larger scattering angles. With decreasing collision
energy the products for each channel are scattered into increasingly
larger scattering angles. A semi-classical surface hopping calculation
can not reproduce the angular dependence of the cross section at
1.7\,eV, because the calculation predicts intensity out to much larger
scattering angles than obtained in this work \cite{nikitin1987}.
Qualitatively, the vibrational level-dependence of the angular cross
section agrees with the model presented by Candori {\it et al.}
\cite{candori2001:jcp}, who obtain the product vibrational levels from
Landau-Zener curve crossing probabilities for vibrationally-adiabatic
intermolecular potentials. They suggest higher $v'$ levels to be
scattered into larger scattering angles, since the time the reaction
complex spends at a particular curve crossing is increased with
decreasing relative energy and thus decreasing impact parameter. In
contrast to previous experiments \cite{Birkinshaw1987:cp} we do not
see an enhancement of higher scattering angles at a particular
relative energy and no backward scattered product ions. That
implies that a scattering resonance is not likely to play an important
role in this reaction system.

\begin{figure}[tb]
  \begin{center}
    \includegraphics[width=1.0\columnwidth, trim=9cm 7cm 2.5cm 1cm, clip]{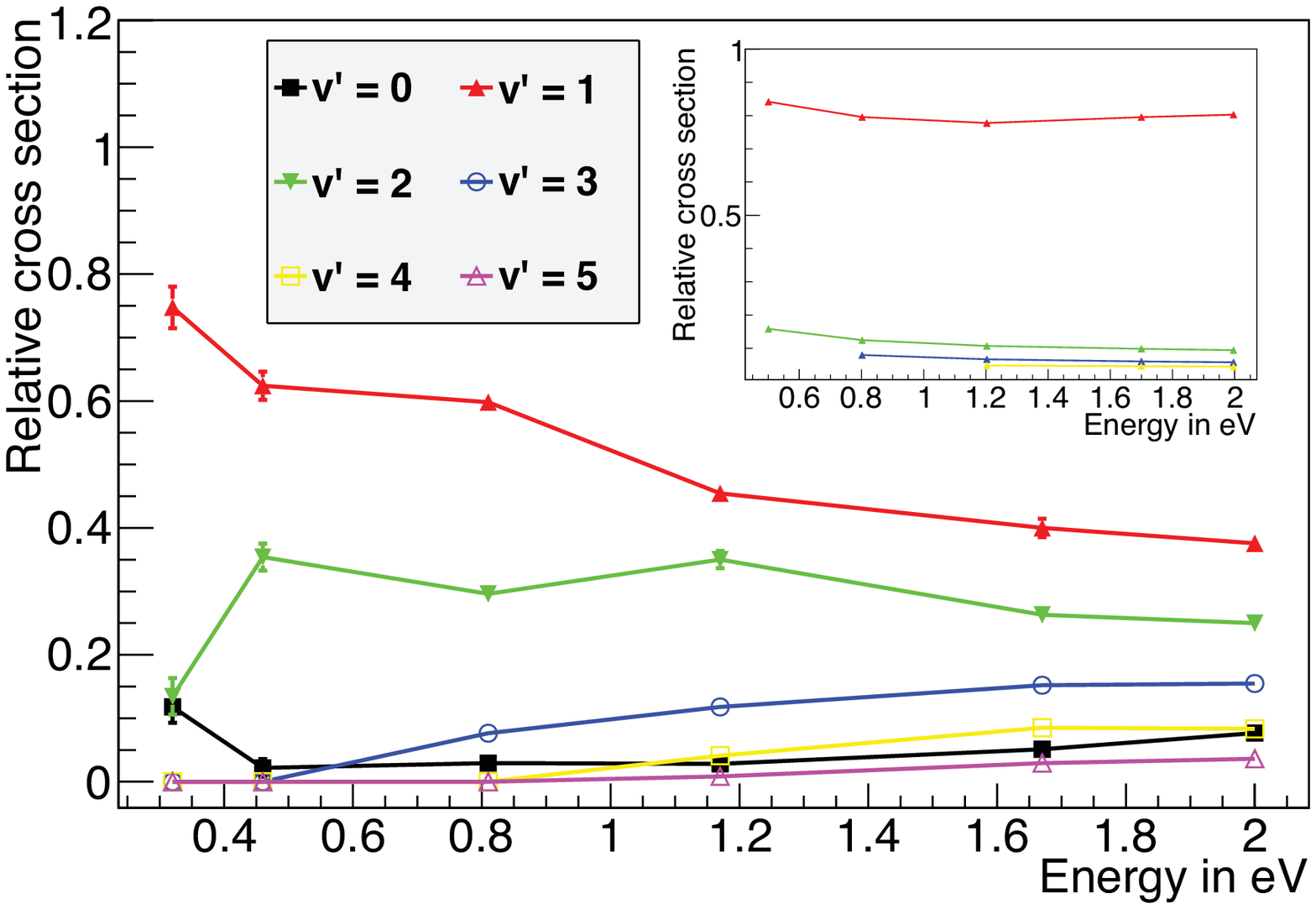}
    \caption{The total product vibrational state-dependent cross
      section as a function of the relative collision energy for
      $v'=0-5$. The inset shows the corresponding theoretical relative
      populations of vibrational states taken from
      Candori\,\cite{candori2001:jcp}.}
    \label{vib_dist_tot.fig}
  \end{center}
\end{figure}

%total vibrational distribution
Fig.\ \ref{vib_dist_tot.fig} presents the vibrational branching ratios
for $v'=0$ to 5, after integration over all scattering angles as a
function of the collision energy. The values are normalized to unity
for each energy. These integral data can be compared to previous
studies of this reaction system. In contrast to the calculation by
Candori {\it et al.}  \cite{candori2001:jcp} (which is shown in the
inset) we find higher vibrational excitation ($v' \ge 2$) to be more
likely. While the calculation yields a ratio of $(v'=2)/(v'=1)$ of
about $1/10$ for all scattering energies, we obtain a ratio of about
$1/2$. This might be attributed to their employed ad hoc-parameters
for the coupling of the vibrational states, or may be caused by the
semi-classical treatment of the problem. Liao {\it et al.}
\cite{liao1986:jcp} measured the product vibrational excitation using
chemical probing and found a ratio of $(v'=2)/(v'=1)$ of about $1/6$
at a collision energy of 1.2\,eV. This is a significantly lower value
compared to our measurement, which might be explained by a
scattering-angle dependent acceptance probability in their experiment.
 
%%%%%%%%%%%%%%%%%%%%%%%%%%%%%%%%%%%%%%%%%%%%%%%%%%%%%%%%%%%%%%%%%%%%%%%%%%%%%%

%Conclusion
In conclusion, we have investigated the charge transfer reaction
Ar$^+$+N$_2\rightarrow$ Ar + N$_2⁺$ between 0.3 and 2\,eV by the
combination of crossed beam techniques and 3D velocity map imaging. We
have presented the detailed energy- and angle-differential cross
sections for a range of collision energies. The achieved experimental
resolution for ion-molecule reactive scattering has been improved
significantly and gets closer the resolution obtained in
neutral-neutral reactions \cite{qiu2006:sci}. The vibrational state
distributions show a clear increase in excitation for larger
scattering angles. This increase becomes more prominent with
decreasing collision energy for all kinematically allowed quantum
states. These observations are qualitatively in line with calculations
\cite{candori2001:jcp}. With some improvements in the calculations, a
full quantitative agreement between experiment and theory, known for
the neutral H + H$_2$ \cite{harich2002:nat} and F + H$_2$
\cite{qiu2006:sci} reactions, now comes within reach for ion-molecule
reactions as well.

This research has been supported by the Deutsche
Forschungsgemeinschaft under contract No. WE 2592/3-2 and by the EU
Marie Curie Initial Training Network {\em ICONIC}. We thank the
University of Freiburg, where the measurements presented here have
been carried out, for supporting this research.

\end{document}